\documentstyle[aps,amsfonts,epsf%
,titlepage%
,floats%
,12pt                       
]{revtex}

\def\wh#1{\widehat{#1}}
\def\cnj#1{\overline{#1}}

\def\union{\bigcup}
\def\inter{\bigcap}

\def\Re{{\rm Re}}
\def\Im{{\rm Im}}

\def\Hin{{\cal H}_{\rm in}}
\def\Hout{{\cal H}_{\rm out}}
\def\inn{{\rm in}}
\def\out{{\rm out}}

\def\qed{{\hfill\vrule height 0.7em width 0.7em depth 0.0em}}
\def\KGspace#1{{{#1}\oplus \cnj{#1}}}
\def\norm{{\cal N}}
\def\suchthat{{\ | \ }}

\def\Cmpx{{\Bbb C}}
\def\Real{{\Bbb R}}
\def\Intg{{\Bbb Z}}

\def\sfrac#1#2{{\displaystyle{\frac{\mbox{\small$#1$ }}
{\mbox{\small$#2$}}}}} 

\newtheorem{lemma}{Lemma}
\newenvironment{proof}
{\par{\bf \noindent Proof}\par}{\qed}

\def\Hin{{\cal  H}(C^\inn)}
\def\Hout{{\cal  H}(C^\out)}
\def\HC{{\cal  H}(C)}

\title%
{ Classical and Quantum Implications
 of the \\
  Causality
Structure of  Two-Dimensional Spacetimes with  Degenerate Metrics \vskip 2em}

\author{
Jonathan Gratus \thanks{gratus{\rm @}ccr.jussieu.fr}\\
School of Physics and Chemistry,
 Lancaster University, \\
Bailrigg, Lancs. LA1 4YB, UK\\
and\\
Laboratoire de Gravitation et Cosmologie Relativistes,\\
Universit\'e Pierre et Marie Curie, CNRS/URA 769\\
Tour 22-12 BP142, 4 Pl. Jussieu, 75252 Paris, Cedex 05\\
\vskip 2em
Robin W Tucker \thanks{r.tucker{\rm @}lancaster.ac.uk} \\
School of Physics and Chemistry,
 Lancaster University, \\
Bailrigg, Lancs. LA1 4YB, UK}

\begin{document}

\maketitle

\vskip 2em

\centerline{\bf Abstract}
\begin{abstract}

The causality structure of two-dimensional manifolds with degenerate
metrics is analysed in terms of global solutions of the massless wave
equation.  Certain novel features emerge.  Despite the absence of a
traditional Lorentzian Cauchy surface on manifolds with a Euclidean
domain it is possible to uniquely determine a global solution (if it
exists), satisfying well defined matching conditions at the degeneracy
curve, from Cauchy data on certain spacelike curves in the Lorentzian
region.  In general, however, no global solution satisfying such
matching conditions will be consistent with this data.  Attention is
drawn to a number of obstructions that arise prohibiting the
construction of a bounded operator connecting asymptotic single
particle states.  The implications of these results for the existence
of a unitary quantum field theory are discussed.
\end{abstract}

\section{Introduction}

If $S$ is a spacelike (acausal) domain in a spacetime $M$ then one
defines its domain of dependence $D(S)$ as the set of points $p$ such
that all non-terminating time-like curves from $p$ intersect $S$
\cite{Geroch1}. Furthermore one says that an acausal hypersurface $S$ is
a Cauchy hypersurface if $D(S)=M$.  The existence of a Cauchy
hypersurface is equivalent to $M$ being globally hyperbolic.

Kundt first {}\cite{Kundt1} discussed the non-existence of certain
topologically non-trivial spacetimes assuming that every geodesic is
complete. Geroch {}\cite{Geroch1} exploited the notion of global
hyperbolicity to reach a similar conclusion.  There has been recent
interest in the behaviour of both classical and quantum fields on
background manifolds that admit metrics with both Euclidean
{}\cite{Gibbons1} and variable signatures 
\cite{Geroch1,Kundt1,Ellis1,%
Dereli1,Dereli2,Dereli3,Hayward1,Kossowski1,Kossowski2,Kossowski3,%
Kerner1,Dray1,Dray2,Jg2,Alty1} 
as well as in spacetimes that are not globally hyperbolic
{}\cite{Kay1}.

For information that propagates according to the wave equation for a
scalar field a specification of the field and its normal derivative on
any spacelike surface $S$ is sufficient to determine a unique solution
to the equation on the domain of dependence of $S$.  A manifold with a
degenerate metric is not globally hyperbolic and it is therefore of
interest to investigate the influence of this degeneracy on the
propagation of massless scalar fields satisfying
\begin{eqnarray}
d\star d\psi=0
\label{LapW} 
\end{eqnarray}
where $\star$ is the Hodge map associated with an ambient metric tensor
field $g$.

In this article we address this question in the context of a metric
that partitions a two-dimensional manifold into three disjoint sets: a
Lorentzian region $L$, a Euclidean region $E$ and a one-dimensional
subset $\Sigma$ where the metric is degenerate.

In a region where $g$ has Euclidean signature (\ref{LapW}) is the
(elliptic) Laplace equation.  The traditional data for this equation
is the specification of $\psi$ on the boundary of any Euclidean domain
since this will fix a solution uniquely in such a region. However we
show below that a specification of $\psi$ and its normal derivative on
any arc of the boundary is also sufficient to uniquely determine the
interior solution should it exist.  This result proves of relevance
when we discuss the propagation of hyperbolic data from a Lorentzian
to a Euclidean domain.

\begin{figure}
\epsfysize=3.5in
\epsffile{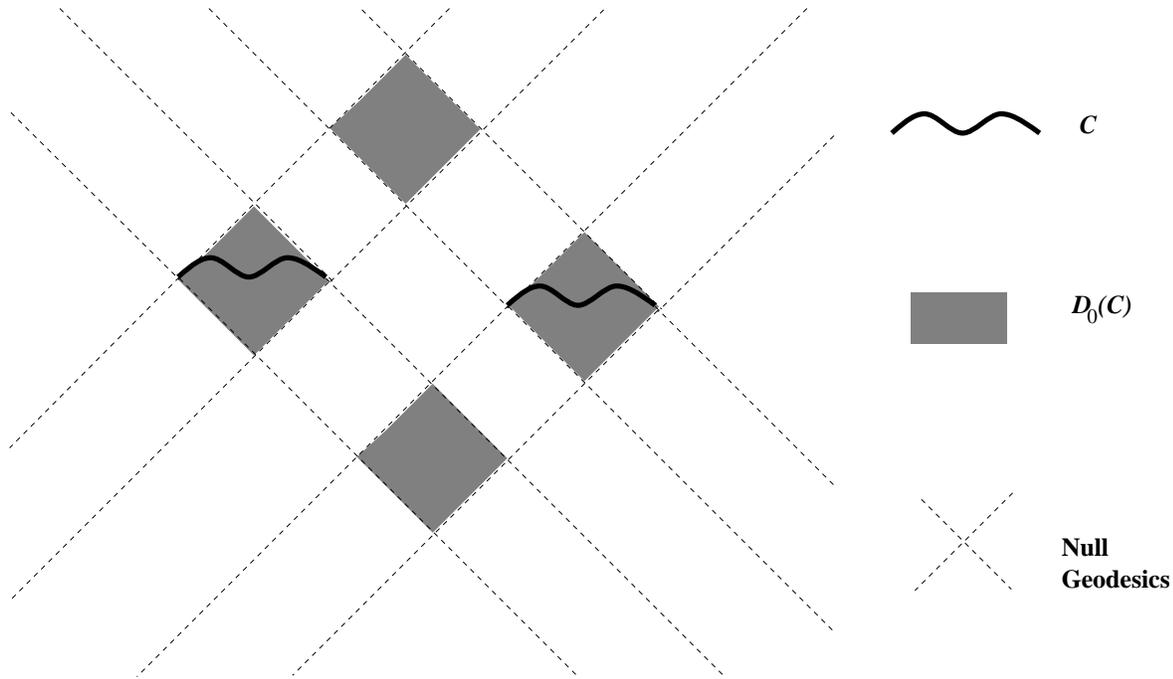}
\caption{{\bf Domain of Null Dependence} of $C$,
$D_0(C)\subset L\union \Sigma$ }
\label{fig_D_0S}
\end{figure}

In a region where $g$ has Lorentzian signature (\ref{LapW}) is the
(hyperbolic) {\it massless} wave equation and it is possible to
contemplate data on disjoint spacelike curves (FIG. \ref{fig_D_0S}).

Since information travels along null geodesics we redefine the domain
of dependence. Thus if $C$ is a spacelike (acausal) domain in a
Lorentzian region $L\subset M$ then its {\it domain of null
dependence} $D_0(C)$ is the set of points $p$ such that all
non-terminating null curves from $p$ intersect $C$.  For $C\subset L$
and $\Sigma'\subset\Sigma$ we define $D_0(C\union\Sigma')$ to be all
the points $p\in L$ where both null geodesics that intersect at $p$,
either intersect $C$ or terminate on $\Sigma'$, where they are not
tangent.

From FIG. \ref{fig_D_0S} we see
that if $C$ is not connected then $D_0(C)$ may contain regions
disjoint from $C$.

\begin{figure}
\epsfysize=6in
\epsffile{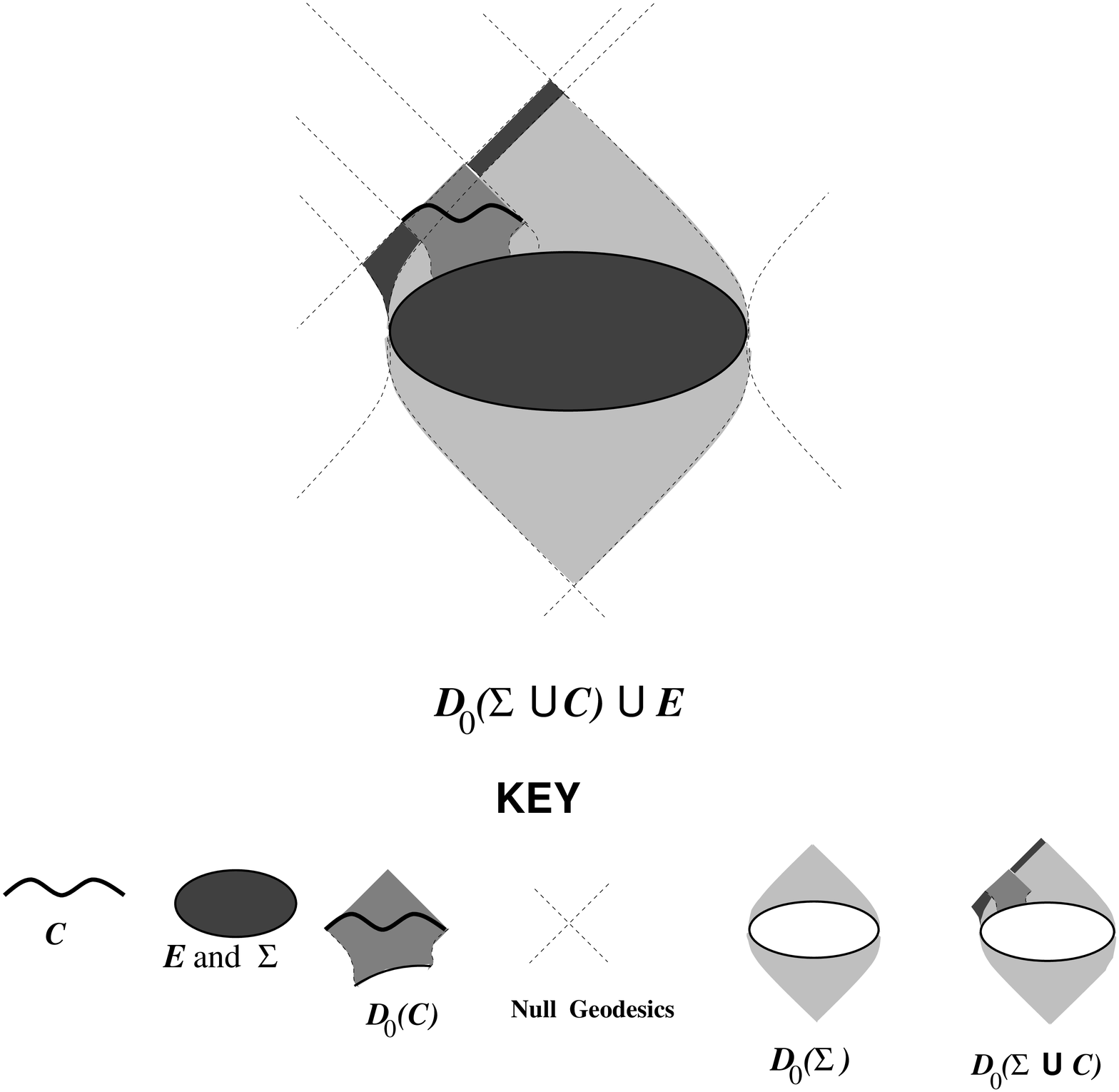}
\caption{$D_0(C\union\Sigma)\union E$ }
\label{fig_DomSandE}
\end{figure} 

Consider now a spacelike arc $C\subset L$ on which standard Cauchy
data for (\ref{LapW}) is prescribed (FIG.
\ref{fig_DomSandE}). Furthermore suppose that the domain of null
dependence of $C$ intersects $\Sigma$ non-trivially, i.e.  the
intersection is one-dimensional.  Then we show below that if a global
solution exists and agrees with the data on $C$ then it is unique in
$E$. Such a solution provides Cauchy data on $\Sigma$ which together
with that on $C$ enables one to construct the solution on the domain
of null dependence $D_0(C\union \Sigma)$.  In general no such global
solution exists as will be illustrated in example 1 below.

If the intersection $D_0(C)\inter\Sigma$ is trivial then the theorem
below is not applicable.  In this case more than one global solution
may exist compatible with the Cauchy data on $C$.  Example 2
will illustrate this situation.

In section \ref{quantisation} we review the standard prescription that
is adopted to  construct a quantum  field theory on  a fixed  globally
hyperbolic  manifold.  With  the aid   of  explicit  examples, we draw
attention to the    obstructions  that arise   when   one attempts  to
implement this  prescription  on  a  two-dimensional manifold  with  a
degenerate   metric.  We offer reasons why   we believe that a bounded
unitary  scattering matrix in the  presence of non-dynamical signature
change may not exist.


\section{Construction of coordinate systems}

Given a two-dimensional manifold $(L,g^L)$ with a (non-degenerate)
Lorentzian metric $g^L$, a {\em null coordinate system}
$(\eta_+,\eta_-)$ is one in which the metric may be written
\begin{eqnarray}
g^L &=& \Omega^L(\eta_+,\eta_-) 
( d\eta_+\otimes d\eta_- + d\eta_-\otimes d\eta_+ )
\label{coord_null_g}
\end{eqnarray}
where $\Omega^L$ is a real function.
Similarly for a two-dimensional manifold $(E,g^E)$ with a
(non-degenerate) Euclidean metric $g^E$, a {\em complex isothermal
coordinate system} $(z, \cnj z)$ is one in which the metric may
written
\begin{eqnarray}
g^E &=& \Omega^E(z,\cnj z)
( dz\otimes d \cnj z + dz\otimes d\cnj z )
\label{coord_cmpx_g}
\end{eqnarray}
where $\Omega^E$ is a real positive function.
It is known that one can construct a null coordinate system about any
point in $p\in L$ and that one can construct a complex isothermal
coordinates system about any point $p\in E$.

In this section we give conditions for the construction of null and
complex isothermal coordinate systems for a two-dimensional manifold
$M=L\union\Sigma\union E$
with a signature changing metric $g$. 

We assume that $\Sigma$ is (locally)
parameterised by a $C^\omega$ monotonic function
$\Sigma:I\mapsto\Sigma$, with $\sigma\in I\subset\Real$. (The symbol
$\Sigma$ will be used for both the map and its image.) 
Given $p\in\Sigma$, we say $\Sigma$ is {\em null at $p$} if
\begin{eqnarray}
g\left.\left(
\Sigma_\star\frac{\partial}{\partial\sigma} \ ,\ 
\Sigma_\star\frac{\partial}{\partial\sigma} 
\right)\right|_p &=& 0
\ .
\label{def_null_Sig}
\end{eqnarray}
If $\Sigma$ is null at $p$ then null geodesics in $L\union \Sigma$ are
tangent to $\Sigma$ at $p$.

For $p\in\Sigma$ such that $\Sigma$ is not null at $p$ Lemma
\ref{lm_extend_null} below shows that there exists a coordinate system
about $p$ in $L$ which can be extended to $\Sigma$, and Lemma
\ref{lm_extend_cmplx} below shows that there exists a complex
isothermal coordinate system about $p$ that extends to $\Sigma$.

Lemma \ref{lm_extend_cmplx} is a modification of the standard proof
for the existence of isothermal coordinates \cite[page 455-460, vol
IV]{Spivak1}. However to simplify matters we restrict our attention to
the case where $g$ is analytic and can be written in {\em absolute
time} form \cite{Kossowski1} about a point $p\in\Sigma$; that is there
exist coordinates $(t,x)$ about $p$ such that the metric can be written
\begin{eqnarray}
g = t\, dt\otimes dt + h(t,x)^2 dx\otimes dx
\label{coord_koss_g}
\end{eqnarray}
where  $h$ is a real positive $C^\omega$ function.  
In this coordinate system the curve $\Sigma$ is given by $t=0$.
Then $g^E$ is given by the restriction 
\begin{eqnarray}
g^E=g\vert_{t>0}
\label{coord_gE}
\end{eqnarray}
From \cite[theorem
1]{Kossowski1} we know that sufficient conditions for writing $g$ in
absolute time form about $p\in\Sigma$ are:

(i) $\Sigma$ is not null at $p$

(ii) with respect to any coordinate system, $\det(g_{ab})$ has a
non-zero differential at $p$, and

(iii) $\{g_{ab}\}$ is $C^\omega$ in a neighbourhood of $p$.

With respect to both the null and complex isothermal
 coordinate system we then derive the general
solution of (\ref{LapW}), specifying both its value and {\em normal
derivative} on $\Sigma$. This data is used to match the solutions
across $\Sigma$.


\begin{lemma} 
{\bf Extension of Null coordinate system.} 
\label{lm_extend_null}
Given $p\in\Sigma$ such that $\Sigma$ is not null at $p$,
there exists a neighbourhood of $p$,
$U^L\subset L\union \Sigma$ and a $C^0$ function
\begin{eqnarray}
\Phi^L : U^L\mapsto 
{\{(\eta_+,\eta_-)\in\Real^2\suchthat\eta_-\le\eta_+\}}
\end{eqnarray}
such that the restriction
\begin{eqnarray}
\Phi^L|_{L} : U^L\inter L\mapsto 
{\{(\eta_+,\eta_-)\in\Real^2\suchthat\eta_- < \eta_+\}}
\end{eqnarray}
is a $C^\infty$ diffeomorphism and a null coordinate system. 
Also
\begin{eqnarray}
\Phi^L|_{\Sigma} &:& U^L\inter \Sigma\mapsto 
{\{(\eta_+,\eta_-)\in\Real^2\suchthat\eta_- = \eta_+\}}
\end{eqnarray}
where
\begin{eqnarray}
\eta_+\circ\Sigma(\sigma)
&=&
\eta_-\circ\Sigma(\sigma)
=
\sigma
\end{eqnarray}
\end{lemma}

\begin{figure}
\epsfysize=2.5in
\centerline{\epsffile{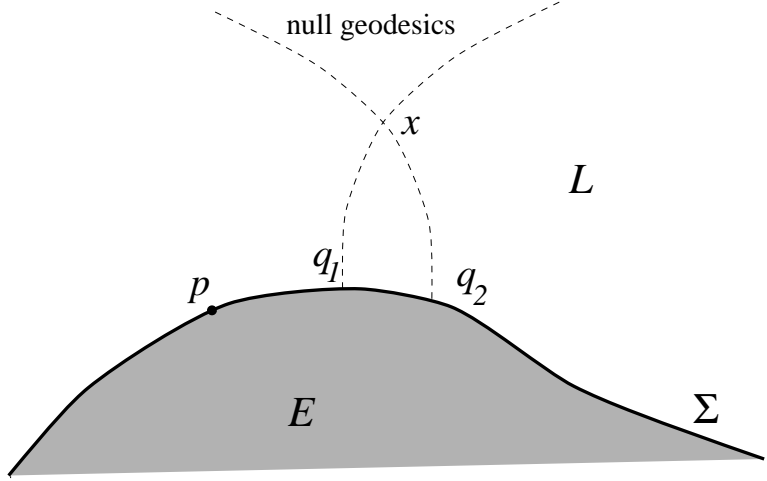}}
\caption{{
Intersection of the null geodesics from $x$ with $\Sigma$ at $q_1$ and
$q_2$.
}}
\label{lm_null_coord_fig}
\end{figure}

\begin{proof}
Every point $x\in L$ near $p$, lies on the intersection of two null
geodesics. We can choose $U^L$ such that for all $x\in U^L\inter L$,
both the null geodesics that intersect $x$ also intersect $\Sigma$
near $p$ where they are not tangent. Thus for each $x$ this gives two
points, $q_1,q_2\in\Sigma$. See FIG. \ref{lm_null_coord_fig}.  Let
$\eta_+(x)$ be the larger of $\{\Sigma^{-1}(q_1),\Sigma^{-1}(q_2)\}$
and $\eta_-(x)$ the smaller.  This construction gives a well defined
$C^\infty$ map $\Phi^L$, which is a diffeomorphism in the set
described.  Furthermore since $\eta_\pm={\rm constant}$ are null
geodesics, $\partial/\partial \eta_\pm$ are null, so the metric is
given by (\ref{coord_null_g}).
\end{proof}

\vskip 1em

With respect to a null coordinate system the Lorentzian Hodge maps
are given by
\begin{eqnarray}
\star 1 &=&
\Omega^L  d\eta_+ \wedge d\eta_- \ ,
\\
\star d{\eta_+} &=& -d{\eta_+} 
\\
\star d\eta_- &=& d\eta_- 
\label{coord_star_eta}
\end{eqnarray}
so the wave equation (\ref{LapW}) for $\psi^L:U^L\to{\Bbb C}$ is
\begin{eqnarray}
\sfrac{\partial^2\psi^L}{\partial {\eta_+}\partial \eta_-}  
d{\eta_+} \wedge d\eta_- = 0
\ ,
\end{eqnarray}
and its solution in this region is:
\begin{eqnarray}
\psi^L=\psi^L_+({\eta_+})+\psi^L_-(\eta_-)
\label{coord_sol_psiL}
\end{eqnarray}
where $\psi^L_\pm:U^L\to{\Bbb C}$.  
We define the value of the
$\psi^L$ on $\Sigma$ and the normal derivative $\norm\psi^L_\Sigma$
of $\psi^L$ on $\Sigma$ by:
\begin{eqnarray}
\psi^L_\Sigma&:&\Sigma\mapsto{\Bbb C}
\nonumber\\
\psi^L_\Sigma(x)&=&\lim_{y\to x,\,\, y\in L}\psi^L(y)
\label{psiSigL}
\\
\norm\psi^L_\Sigma &:& \Sigma\mapsto{\Bbb C}
\nonumber\\
\norm\psi^L_\Sigma(x) &=& 
i_{\frac{\partial}{\partial \sigma}}
\left( \Sigma^\star \left(
\lim_{y\to x,\,\, y\in L}
\star d\psi^L|_y
\right)
\right)
\label{psiNormL}
\end{eqnarray}
With respect to the coordinate systems $(\eta_+,\eta_-)$ given in lemma
\ref{lm_extend_null}, these are given by;
\begin{eqnarray}
\psi^L_\Sigma(x)&=&\psi^L_+({\eta_+(x)})+\psi^L_-(\eta_-(x))
\\
\norm\psi^L_\Sigma(x)
&=&
i_{\frac{\partial}{\partial \sigma}}
\left(
\Sigma^\star
\lim_{y\to x,\,\, y\in L}
\star d
\left(
\psi^L_+({\eta_+})+{\psi^L_-(\eta_-)}
\right)|_y
\right)
\cr
&=&
i_{\frac{\partial}{\partial \sigma}}
\left(
\Sigma^\star
\lim_{y\to x,\,\, y\in L}
\left(
\psi^L_+{}'\star d\eta_+ +\psi^L_-{}'\star d\eta_-
\right)|_y
\right)
\cr
&=&
i_{\frac{\partial}{\partial \sigma}}
\left(
\Sigma^\star
\lim_{y\to x,\,\, y\in L}
\left(
-\psi^L_+{}' d\eta_+ +\psi^L_-{}' d\eta_-
\right)|_y
\right)
\cr
&=&
i_{\frac{\partial}{\partial \sigma}}
\left(
\Sigma^\star
\left(
-\psi^L_+{}'d\eta_+ +\psi^L_-{}'d\eta_-
\right)|_x
\right)
\cr
&=&
-\psi^L_+{}' \frac{\partial\eta_+\circ\Sigma}{\partial \sigma} 
+\psi^L_-{}' \frac{\partial\eta_-\circ\Sigma}{\partial \sigma}
\cr
&=&
-\psi^L_+{}'(\eta_+(x)) +\psi^L_-{}'(\eta_-(x)) 
\end{eqnarray}
since $\partial \eta_\pm\circ\Sigma / \partial\sigma = 1$. 
Thus the
normal derivative exists so long as  $\Sigma$ is not null.

\vskip 1em


\begin{lemma} 
{\bf Extension of Complex Isothermal coordinate system.} 
\label{lm_extend_cmplx}
Given $p\in\Sigma$  such that $g$ is analytic and can be written in
absolute time form (\ref{coord_koss_g}) about a point $p\in\Sigma$,
then there exists a neighbourhood of $p$, $U^E\subset E\union \Sigma$
and a $C^0$ function
\begin{eqnarray}
\Phi^E : U^E\mapsto 
{\{z\in\Cmpx\suchthat\Im(z)\ge0\}}
\end{eqnarray}
such that the restriction
\begin{eqnarray}
\Phi^E|_{E} : U^E\inter E\mapsto 
{\{z\in\Cmpx\suchthat\Im(z) > 0\}}
\label{PhiE2}
\end{eqnarray}
is a $C^\omega$ diffeomorphism, and a complex isothermal coordinate.
Also
\begin{eqnarray}
\Phi^E|_{\Sigma} &:& U^E\inter \Sigma\mapsto \Real
\label{PhiE3}
\end{eqnarray}
\end{lemma}

\begin{proof}{}
Let $g$ be written as in (\ref{coord_koss_g}), so that  $p$ is given by
$(t=0,x=x_0)$ and  $g^E$ is given by ({\ref{coord_gE}}).
Let
\begin{eqnarray}
\omega &=& h\, dx + i\sqrt{t}\, dt
\end{eqnarray}
so that
\begin{eqnarray}
g^E &=& \omega \otimes \cnj{\omega} + \cnj{\omega} \otimes \omega 
\ .
\label{g_in_om}
\end{eqnarray}
We seek a non-vanishing integrating factor $\lambda(t,x)$, and
a complex coordinate $z(t,x)$ such that
\begin{eqnarray}
\omega\lambda=dz\ .
\label{int_fact}
\end{eqnarray}
Consider the extension of $\omega$ to $\Real^+\times\Cmpx$ given by
\begin{eqnarray}
\omega &=& h(t,\xi)\, d\xi + i\sqrt{t}\, dt
\end{eqnarray}
where $t\in\Real, t>0$ and $\xi\in\Cmpx$, and where $h(t,\xi)$ is the
analytic extension of $h(t,x)$ with respect to its second argument.
Let
\begin{eqnarray}
X &=& 
\frac{-i\sqrt{t}}{h} \frac{\partial}{\partial\xi}
+
\frac{\partial}{\partial t}
\ ,
\end{eqnarray}
a vector field on $\Real^+\times\Cmpx$, which is an
annihilator of $\omega$, i.e. $\omega(X)=0$. The solution curves of $X$
are the set of curves $t\mapsto (t,\xi(t))$, where $\xi(t)$ are solutions
to the differential equation:
\begin{eqnarray}
\frac{dx}{dt}=-i\frac{\sqrt{t}}{h(x,t)}
\ .
\label{cause_cmpx_coord_ODE}
\end{eqnarray}
We note that the component of $X$ in the $t$
direction is 1, and the component of $X$ in the
imaginary $\xi$ direction is
\begin{eqnarray}
-\frac{\sqrt{t}\,\Re(h)}{|h|^2} 
\end{eqnarray}
%
%
\begin{figure}
\epsfysize=2.5in
\centerline{\epsffile{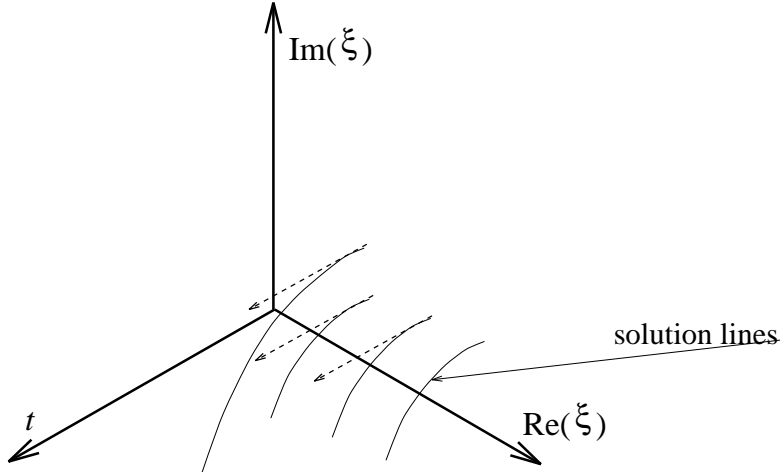}}
\caption{{ The solution lines of $X$ intersecting with the half plane
$\{\Im(\xi)=0,t>0\}$ and the half plane $\{t=0,\Im(\xi)>0\}$ where
they are parallel to the field $\partial/\partial t$.  }}
\label{lm_cmpx_coord_fig}
\end{figure} 
which, near $(t=0,\xi=x_0)$ is bounded above by $\sqrt{t}$ times some
factor. Therefore near $p=(0,x_0)$, all solution curves of
(\ref{cause_cmpx_coord_ODE}) that intersect the half plane
$\{t=0,\Im(\xi)>0\}$ also intersect the plane $\{\Im(\xi)=0,t>0\}$ and
visa versa. See FIG. \ref{lm_cmpx_coord_fig}. Thus we have the map
\begin{eqnarray}
\Phi^E &:& U^E\mapsto 
{\{z\in\Cmpx\suchthat\Im(z)\ge0\}}
\cr
\Phi^E &:& (t,x) \mapsto z
\end{eqnarray}
where the solution curve connects the point
$(t=t,\Re(\xi)=x,\Im(\xi)=0)\in\{\Im(\xi)=0,t>0\}$ to the point
$(t=0,\xi=z)\in\{t=0,\Im(\xi)>0\}$.  This map satisfies (\ref{PhiE2}),
and (\ref{PhiE3}).  Since $z$ now labels each solution curve, and is
thus a constant along it; $X(z)\equiv dz(X)=0$.  In general
\begin{eqnarray}
dz &=& \lambda \omega + \mu_1 \cnj{\omega} + \mu_2 dt
\end{eqnarray}
but contracting with $\partial/\partial\cnj{\xi}$ implies that
$\mu_1=0$ since $z$ and $\omega$ are analytic. Contracting with $X$
gives $\mu_2=0$ so (\ref{int_fact}) holds.

Substituting $dz$ into (\ref{g_in_om}) gives (\ref{coord_cmpx_g}) with
$\Omega^E=1/\lambda\cnj{\lambda}$.
\end{proof}

\vskip 1em

The Euclidean Hodge maps are given with respect to the $(z,\cnj z)$
coordinate system by
\begin{eqnarray}
\star1 &=& \Omega^2 dz \wedge d\cnj z
\\
\star d{z} &=& -i\;d{z} 
\\
\star d\cnj{z} &=& i\;d\cnj{z} 
\label{coord_star_z}
\end{eqnarray}
so Laplace's equation for $\psi^E:U\to{\Bbb C}$ is
\begin{eqnarray}
\sfrac{\partial^2\psi^E}{\partial {z}\partial \cnj{z}}  
d{z} \wedge d\cnj{z} = 0
\end{eqnarray}
and its solution in this region is
\begin{eqnarray}
\psi^E=\psi^E_+({z})+\cnj{\psi^E_-(z)}
\ ,
\label{coord_sol_psiE}
\end{eqnarray}
where $\psi^E_\pm:U\to{\Bbb C}$. Since $\partial\psi^E_\pm/\partial
\cnj{z}=0$ these are analytic functions.  We define $\psi^E_\Sigma$
and $\norm\psi^E_\Sigma$ by replacing the symbol $L$ with $E$ in
(\ref{psiSigL}) and (\ref{psiNormL}).  These are given by;
\begin{eqnarray}
\psi^E_\Sigma(p)&=&\psi^E_+(z(p))+\cnj{\psi^E_-(z(p))}
\label{psiESig}
\\
\norm\psi^E_\Sigma(p)
&=&
-i\psi^E_+{}' \frac{\partial z}{\partial \sigma} 
+i\cnj{\psi^E_-{}'} \frac{\partial \cnj z}{\partial \sigma}
\end{eqnarray}

We have
\begin{eqnarray}
\frac{\partial z}{\partial \sigma} 
&=&
\frac{dx}{d\sigma}
\quad \in\Real-\{0\}
\end{eqnarray}
hence
\begin{eqnarray}
\norm\psi^E_\Sigma(p)
&=&
i \frac{dx}{d\sigma}
\left(
-\psi^E_+{}' + \cnj{\psi^E_-{}'}
\right)
\ .
\label{NormpsiESig2}
\end{eqnarray}

So the normal
derivative exists about the point $p\in\Sigma$ so long as $\Sigma$ is
not null.


\section{Uniqueness Lemma}

\begin{lemma}
\label{lm_uniqE}
Let $(E\union\Sigma,g)$ be a closed pathwise connected manifold with
boundary $\Sigma=\partial E$, with a Riemannian metric $g$ which is
degenerate on $\Sigma$ and analytic in a neighbourhood of $\Sigma$.
Let there be only isolated points on $\Sigma$ about which $g$ cannot
be written in absolute time form.  Given a non-trivial curve
$C:I\mapsto\Sigma$ where $I$ is some interval in ${\Bbb R}$
parameterised by $s$ and two functions $u^E,v^E:E\mapsto{\Bbb C}$ that
satisfy Laplace's equation $d\star d\psi=0$ on the interior of $E$,
then $u^E_\Sigma|_C=v^E_\Sigma|_C$ and $\norm u^E_\Sigma|_C=\norm
v^E_\Sigma|_C$ if and only if $u^E=v^E$.

In these equations $u^E_\Sigma(x),\,\norm u^E_\Sigma(x)$ are defined
by replacing $\psi$ by the symbol $u$ in the Euclidean version of
(\ref{psiSigL}),(\ref{psiNormL}).
\end{lemma}

\begin{proof}

Let $(U^E,\Psi^E)$ be a complex isothermal chart of $E$
about a point $p\in C(I)$ given by lemma \ref{lm_extend_cmplx} above.
For the solution $w^E=u^E-v^E$, if follows that $w^E_\Sigma|_C=0$ and
$\norm w^E_\Sigma|_C=0$, thus from (\ref{psiESig}) and
(\ref{NormpsiESig2}) we have
\begin{eqnarray}
w^E_\pm(z) &=& 0
\qquad \forall z\in \Psi^E\circ C(I)
\end{eqnarray}
hence $w^E_\pm$ are analytic functions on the domain $\Psi^E(U^E\inter
E)\subset\Cmpx$ where $\Im(\Psi^E(U^E\inter E))> 0$. They are
continuous functions on the domain $\Psi^E(U^E)\subset\Cmpx$ where
$\Im(\Psi^E(U^E\inter\Sigma))= 0$.  One can perform a Schwartz
reflection \cite{Lang1} about $\Im(z)=0$ to produce an analytic
function on a domain with a non trivial subset of $\Psi^E(\Sigma)$ in
its interior. Since $\Psi^E(C)$ is not a distinct collection of points
\begin{eqnarray}
w^E_\pm(z) &=& 0
\qquad \forall z\in \Psi^E(U^E)
\ .
\end{eqnarray}

For any other chart $(\tilde U,\tilde\Psi)$ where $U\inter\tilde
U\ne\emptyset$, since
\begin{eqnarray}
w^E_\pm(z)=0 ,\qquad\forall z\in \tilde\Psi(\tilde U\inter U) 
\end{eqnarray}
then
\begin{eqnarray}
w^E_\pm(z)=0 ,\qquad\forall z\in \tilde\Psi(\tilde U) \ .
\end{eqnarray}
Thus $w^E_\pm=0$, or $u^E=v^E$ on $E$.
\end{proof}

\vskip 1em


This is a uniqueness proof not an existence proof.  Given two
functions $f_1,f_2:C(I)\mapsto{\Bbb C}$ there will in general be no
solution $u:E\mapsto{\Bbb C}$ to Laplace's equation such that
$u|_C=f_1$ and $i_{C_\star \partial / \partial s }\star u=f_2$ for any
neighbourhood of $C(I)$.
\vskip 1em

{\bf Example 1} \nobreak 

Let $\Psi^E(U^E)=\{z\in\Cmpx\suchthat\Im(z)<0\}$ and
$\Psi^E(C)=\{z\in\Cmpx\suchthat\Im(z)=0\}$. 
Let
\begin{eqnarray}
\psi^E_\Sigma(x) &=&
\left\{
\matrix{
\displaystyle{e^{({-1/x^2})}} \qquad & x > 0 \cr
0 & x \le 0
}\right.
\cr
\norm\psi^E_\Sigma(x) &=&
0 \qquad\qquad \forall x\in\Real
\label{non_anal_data}
\end{eqnarray}
where $x\in\Psi^E(C)$.  According to Lemma \ref{lm_uniqE}, the data on
$x<0$ implies that $\psi(z)=0\quad\forall z\in\Psi^E(U^E)$, whilst the
data on $x>0$ implies that $\psi(z)=\exp(-1/x^2) \quad\forall
z\in\Psi^E(U^E)$.  Thus there is no solution to Laplace's equation on
$\Psi^E(U^E)$ consistent with this boundary data.

This does not contradict the Cauchy-Kowalewski theorem
\cite{Mizo1,Courant1} on the existence of solutions to PDE's with
analytic Cauchy data since the data (\ref{non_anal_data}) is not
analytic.

\vskip 1em

\section{Uniqueness Theorem}

In order to match Euclidean solutions to Lorentzian solutions, we must
adopt some boundary conditions along the degeneracy curve $\Sigma$.

The class of boundary conditions adopted here relate boundary data
in a {\bf linear} and {\bf pointwise invertible} manner. Thus for
$J:\Sigma\mapsto GL_2({\Bbb C})$
\begin{eqnarray}
\pmatrix{\psi^L_\Sigma(x) \cr \norm\psi^L_\Sigma(x)}
=
J(x)
\pmatrix{\psi^E_\Sigma(x) \cr \norm\psi^E_\Sigma(x)}
\label{junct}\end{eqnarray}

The ``natural'' boundary conditions given by continuity of $\psi$ and
its normal derivative across $\Sigma$ correspond to 
\begin{eqnarray}
J(x)={\bf 1}.
\label{natjunc}
\end{eqnarray}
We note in passing that the theorem below is applicable even when
$\psi$ is restricted such that
$\norm\psi^E_\Sigma(x)=\norm\psi^L_\Sigma(x)=0$.
\vskip 2em

{\bf Theorem}
\begin{it}
Let (M,g) be a two-dimensional manifold $M$ with a metric $g$ that is
degenerate along a curve $\Sigma$, partitioning $M$ into a Lorentzian
domain $L$ and a connected Euclidean domain $E$. Let $g$ be analytic
in a neighbourhood of $\Sigma$, and let there only be isolated points
on $\Sigma$ where $g$ cannot be written in absolute time form.  Let
$C\subset L$ be an acausal curve, parameterised by $s$, such that
$D_0(C)\inter\Sigma$ contains an arc.  Given two solutions of
(\ref{LapW}) $u,v:M\mapsto{\Bbb C}$ satisfying any boundary condition
in the class above, then $u|_C=v|_C$ and $(C^\star \star du)(\partial/
\partial s) = (C^\star \star dv)(\partial/ \partial s)$ if and only if
$u|_{D_0(C\union\Sigma)\union E}=v|_{D_0(C\union\Sigma)\union E}$ i.e.
$u$ and $v$ agree on the entire shaded area indicated in FIG.
\ref{fig_DomSandE}.
\end{it}

\begin{proof}

Since $u$ and $v$ have the same Cauchy data on $C$, then if $w=u-v$ it
follows that $w|_{D_0(C)}=0$. Hence $w$ has zero Cauchy data on the
Lorentzian side of $D_0(C)\inter\Sigma$ i.e.
\begin{eqnarray}
w^L_\Sigma(x)=0
\ \mbox{and}\  
\norm w^L_\Sigma(x)=0
\qquad \forall x\in D_0(C)\inter\Sigma\ .
\end{eqnarray}
In these equations $w^L_\Sigma(x),\,{\cal
N}w^L_\Sigma(x),\,w^E_\Sigma(x),\,\norm w^E_\Sigma(x)$ are defined by
replacing $\psi$ by $w$ in (\ref{psiSigL}),(\ref{psiNormL}).  Since
the boundary conditions are linear and pointwise invertible we have
zero Cauchy data for $w$ on the Euclidean side of
$D_0(C)\inter\Sigma$.  i.e.
\begin{eqnarray}
w^E_\Sigma(x)=0
\ \mbox{and}\  
\norm w^E_\Sigma(x)=0
\qquad \forall x\in D_0(C)\inter\Sigma
\end{eqnarray}
Hence $w|_E=0$ from Lemma \ref{lm_uniqE}.  This now implies zero Cauchy
data for $w$ on the Euclidean side of the {\it whole} of $\Sigma$.
i.e.
\begin{eqnarray}
w^E_\Sigma(x)=0
\ \mbox{and}\  
\norm w^E_\Sigma(x)=0
\qquad \forall x\in \Sigma
\end{eqnarray}
Thus these boundary conditions give zero Cauchy data for $w$ on the
Lorentzian side of the {\it whole} of $\Sigma$. Since the Cauchy data
for $w$ on $C$ is zero as well, it follows that
$w|_{D_0(C\union\Sigma)\union E}=0$.
\end{proof}

\vskip 1em
 
\begin{figure}
\epsfysize=3in
\epsffile{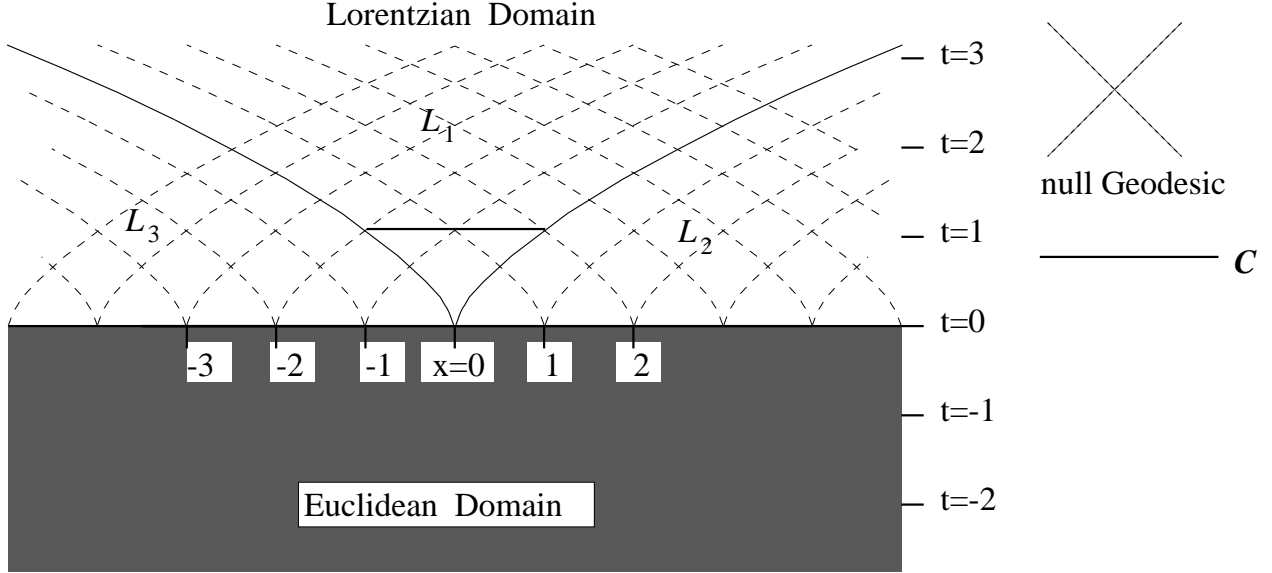}
\caption{Example where $D_0(C)\inter\Sigma$ is trivial }
\label{fig_emptyint}
\end{figure}

Again we note that this is a uniqueness theorem. In general no
solution to (\ref{LapW}) on $M$ will exist satisfying (\ref{junct})
with Cauchy data on some $C$ satisfying the conditions of this
theorem.  The requirement that $D_0(C)\inter\Sigma$ is non-trivial is
necessary.  If the domain of null dependence of $C$ contains a single
point in common with the Euclidean domain $E$ it is not difficult to
construct a situation in which there exists more than one solution on
$E$ compatible with the Cauchy data. 
\vskip 1em

{\bf Example 2} \nobreak 
Let $M={\Bbb R}^2,\ g=-\frac94 t dt\otimes dt + dx\otimes dx$.  We
wish to find solutions that have zero Cauchy data on $C=\{(-1,x),-1\le
x\le 1\}$.  The null geodesics (denoted by dotted lines in FIG.
\ref{fig_emptyint}) are given by $x\pm t^{(3/2)}=\mbox{constant}$.
Clearly one solution that also satisfies the natural boundary
conditions (\ref{natjunc}) is $\psi=0$ on $M$.  However one readily
verifies that another solution is
\begin{eqnarray}
\psi|_L(x,t)&= & \left\{
\vcenter{
\halign{$#\hfil$&$#\hfil$&$#\hfil$&$#\hfil$&$#\hfil$&$#\hfil$&$#\hfil$\cr
0 &
\quad &
x+t^{3/2}>0 \,, &
\, & 
x-t^{3/2}<0 \,, &
\, & 
( L_1 \mbox{ in FIG. \ref{fig_emptyint}})   \cr
(1-i)\sqrt{x-t^{3/2}} & 
&
x-t^{3/2}>0 \,, &
& 
t>0 \,,&
& 
( L_2 \mbox{ in FIG. \ref{fig_emptyint}})   \cr
(1-i)\sqrt{-x-t^{3/2}} & 
&
x+t^{3/2}>0 \,, &
& 
t>0 \,,&
& 
( L_3 \mbox{ in FIG. \ref{fig_emptyint}})   
%
\cr
}}
\right.
\\
\psi|_E(x,t)&=&
\left(\sqrt{x+i t^{3/2}}\right)\ +
\  i\,\cnj{\left(\sqrt{x+i t^{3/2}}\right)}
\end{eqnarray}
where $\sqrt{\strut\ }:\{z\in{\Bbb C}:\Im(z)\le 0\} \mapsto\ 
\{z\in{\Bbb C}:\Re(z)\ge 0,\,\Im(z) \le 0\}$.


\vskip 1em

In a two-dimensional universe with a known degenerate metric one might
imagine one could use the theorem above to predict the behaviour of
the scalar field beyond causally connected regions.  However realistic
Cauchy data that is obtained from physical measurements will contain
errors.  Since it is possible to find global solutions that lie
arbitrarily close to such data in the domain of the Cauchy curve but
are arbitrarily disparate elsewhere one must conclude that the
propagation of such errors cannot be controlled.  \vskip 1em {\bf
Example 3} \nobreak Referring to FIG.~\ref{fig_emptyint}, enlarge $C$ to
$\wh C=\{(t=1,x):|x|<2\}$. Assume the ``experimental'' data given on
$\wh C$, prescribes that a function and its normal derivative are zero
to an accuracy $\epsilon$.  If $\epsilon$ is zero then the theorem
above implies that the only global solution is identically zero.  For
$\epsilon>0$, given any continuous function $f:\Real\mapsto\Cmpx$ such
that $f(x)=0,\ \forall |x|\le 3$, then from the Boltzano-Wiesstrass
theorem there exists a polynomial ${\cal P}:\Real\mapsto\Cmpx$ such
that $|{\cal P}(x)-f(x)|<\epsilon/2,\ \forall x\le5$.  The function
${\cal P}(x+it^{3/2})$ solves Laplace's equation (\ref{LapW}) in the
Euclidean domain and $\vert{\cal P}(x)\vert < \epsilon/2$ on the
interval $\{(t=0,x):|x|<3\}$.  However, this function extends to a
global solution of (\ref{LapW}) on $\Real^2$ and is consistent with
the ``experimental'' Cauchy data on $\wh C$ within the prescribed
error. This is similar to Hadamard's example \cite{Saka1} demonstrating
that certain initial value problems are not {\it well posed} for
Laplace's equation.


\section{Implications for  Quantisation.}
\label{quantisation}

The standard method \cite{Fulling1,Wald1,Wald2} for
constructing a quantum field theory in a curved spacetime is to
consider a globally hyperbolic manifold $M$ possessing Cauchy surfaces
$C^\inn$ and $C^\out$ in (asymptotically) flat regions
upon which one sets up
Hilbert spaces $\KGspace\Hin$ and  $\KGspace\Hout$
 of solutions  to local field equations.

The Klein-Gordon hermitian bilinear form is defined as:
\begin{eqnarray}
<u,v>_{C}=\frac{1}{2\pi i}\int_{C} 
\cnj{u} \star d v -
v \star d \cnj{u} 
\end{eqnarray}
where $u,v$ are  Cauchy data on a Cauchy  surface $C$ for solutions to
the scalar wave (Klein-Gordon) equation. For $C$  in  a flat region of
spacetime the (maximal)  subspace  of  positive frequencies, $\HC$, is
defined by some timelike vector field which is Killing in this region,
such that the restriction of  $<,>_C$ to  $\HC$  is positive definite.
The restriction of $<,>_C$  to its conjugate  $\cnj{\HC}=\{\cnj{u}\ |\
u\in\HC\}$ is negative definite. We have
\begin{eqnarray}
<u,v>_C&=&
<u_+,v_+>_C+<\cnj{u_-},\cnj{v_-}>_C
\end{eqnarray}
where $u\!=\!u_+\!+\!\cnj{u_-}\ ,\ v\!=\!v_+\!+\!\cnj{v_-}$ and
$u_\pm,v_\pm\in\HC$.  The Hilbert space $\KGspace\HC$ is defined with
respect to the ``true'' inner product
\begin{eqnarray}
\ll u,v\gg_C&=&
<u_+,v_+>_C-<\cnj{u_-},\cnj{v_-}>_C\ .
\end{eqnarray}


For any solution $u\!:\!M\mapsto\Cmpx$ there is Cauchy data
associated with $u^\inn\in\KGspace{\Hin}$ for $u$, and Cauchy data
associated with $u^\out\in\KGspace{\Hout}$. The corresponding quantum
system is said to be {\em unitary} if
\hbox{$<u^\inn,v^\inn>_{C^\inn}=<u^\out,v^\out>_{C^\out}$}. This is
guaranteed for a globally hyperbolic manifold since the current
$\cnj{u} \star d u - u \star d \cnj{u}$ is conserved.  The linear map
$\cal A:\KGspace{\Hin} \mapsto \KGspace{\Hout}$ defined by
$u^\inn\mapsto u^\out={\cal A}u^\inn$ may be represented as
\begin{eqnarray}
{\cal A} = \pmatrix{\alpha & \beta \cr \cnj{\beta} & \cnj{\alpha}}
\end{eqnarray}
and defines the Bogolubov transformations:
\begin{eqnarray}
\alpha &:&{\Hin} \mapsto {\Hout}
\\
\beta &:&{\Hin} \mapsto \cnj{\Hout}
\end{eqnarray}

From these one may  construct the Scattering matrix
in a  Fock space basis  of many particle states in the quantum theory.
The expectation value of the particle number density with respect to
the image of a Fock space vacuum state under this Scattering matrix
can be shown to be $\sum_{ij}|\beta_{ij}|^2$. For a finite theory
$\beta$ must be Hilbert-Schmidt.  i.e.
$\sum_{ij}|\beta_{ij}|^2<\infty$ \cite[page 140]{Fulling1}.  This
implies that ${\cal A}$ must be bounded.

\vskip 2em

It is of interested to see to  what extent  the formalism above breaks
down  in  the context of a  quantum  field  theory  of massless scalar
particles in a background two-dimensional  spacetime with a degenerate
metric.  We approach this by considering a  simple example in which we
can  readily  calculate  the  Bogolubov  coefficients.   Let $M$ be  a
cylinder with   coordinates  $\{(\tau,\theta)  |  -\infty<\tau<\infty,
0\le\theta<2\pi\}$, together with the axially symmetric metric
\begin{eqnarray}
g=g_{\tau\tau}(\tau) d\tau\otimes d\tau + 
g_{\theta\theta}(\tau) d\theta\otimes d\theta
\end{eqnarray}
where $g_{\theta\theta}(\tau)>0 $ for all $\tau$.  $M$ has a Euclidean
region $E$ (where $g_{\tau\tau}>0$) sandwiched  between two Lorentzian
regions labelled   $L^\inn$ and  $L^\out$ (where $g_{\tau\tau}<0$). Let
there be  a flat region  $L_{\rm flat}^\inn\subset  L^\inn$ i.e. where
$g_{\tau\tau}=-1,\,g_{\theta\theta}=1$, containing  the Cauchy surface
$C^\inn\subset L_{\rm flat}^\inn$. Similarly let $C^\out\subset L_{\rm
flat}^\out\subset L^\out$.   Let $\Sigma^\inn$ be the  degeneracy ring
$\tau=\tau^\inn$ that partitions $E$ from $L^\inn$.  Similarly suppose
$\Sigma^\out$ is the degeneracy ring $\tau=\tau^\out > \tau^\inn$ that
partitions $E$ from $L^\out$.   See \cite{Jg2} for explicit  details of
such a construction.

A  transformation to complex  coordinates, such that the metric  may be
written as (\ref{coord_cmpx_g}), is given by
\begin{eqnarray}
z=\Psi(\tau,\theta) &=& 
e^{i\theta} \exp\left(\int_{\tau^\inn}^\tau
\left( g_{\tau\tau}(\tau') \over g_{\theta\theta}(\tau') \right)
^\frac12 d\tau' \right)
{\rm \ \ where \ \ }
\tau^\inn\le\tau\le\tau^\out,\,\,\, 0\le\theta<2\pi
\label{defPsi}
\end{eqnarray}
Thus $\Psi(E)$ is an annulus about the origin with radii $1$ and $e^\xi$
where
\begin{eqnarray}
\xi=\int_{\tau^\inn}^{\tau^\out} 
\left( g_{\tau\tau}(\tau') \over g_{\theta\theta}(\tau') \right)
^\frac12 d\tau'\, .
\end{eqnarray}

A convenient basis of non-zero mode 
solutions to (\ref{LapW}) in the  Lorentzian flat domain
 $L_{\rm flat}^\inn$  is
\begin{equation}
\left\{
e^\inn_k=|2\pi k|^{-\frac12} e^{-i(|k|\tau+k\theta)}\ ,\ 
\cnj{e^\inn_k}=|2\pi k|^{-\frac12} e^{i(|k|\tau+k\theta)}
\right\}_{
k\in\Intg,\,\, k\ne 0 }
\end{equation}
where $(\tau,\theta)$ lie in $L_{\rm flat}^\inn$. 
This basis is in 1-1 correspondence with the set of Cauchy data on $C^\inn$
and hence provides a  
basis  of one-particle  in-states
for  $\KGspace\Hin$, with $\{e^\inn_k\}$ giving a basis for $\Hin$. 
A basis
of one-particle out-states  is likewise defined  from
$\{e^\out_k,\cnj{e^\out_k}\}$ where 
$(\tau,\theta)$ lie in $L_{\rm flat}^\out$.


The Bogolubov coefficients can easily be  calculated from the matching
conditions (\ref{natjunc}):
\begin{eqnarray}
\alpha_{kk}&=&\cosh(k\xi)
\nonumber\\
\alpha_{kj}&=&0\qquad j\ne k
\\
\beta_{kk}&=&i\sinh(k\xi)
\nonumber\\
\beta_{kj}&=&0\qquad j\ne k
\label{bogeuclid}
\end{eqnarray}

It is obvious that  $\beta$ is  not Hilbert-Schmidt. Furthermore it is easy to
verify that
the linear map ${\cal A}:\KGspace{\Hin} \mapsto \KGspace{\Hout}$ is not even
bounded. This follows  since one can construct a solution $u$
on $M$ which has a pole on $\Sigma^\out$, e.g. 
\begin{eqnarray}
u(\tau ,\theta)={1 \over \Psi(\tau ,\theta)-\Psi(\tau^\out,0)} 
\end{eqnarray}
Now $\|u^\inn\|_{C^\inn}<\infty$ but $\|u^\out\|_{C^\out}=\infty$,
which could not happen if $\cal A$ were bounded.
($\|{\cal A}(u^\inn)\|_{C^\out}=
\|u^\out\|\le\|{\cal A}\|\,\|u^\inn\|_{C^\inn}$.)
However on the subspace consisting  of  finite sums  of basis elements
above, the mapping $\cal A$ has a unitary restriction.  Furthermore it
can  be shown      that   if   $u:M\mapsto\Cmpx$  is   bounded,    and
$u^\inn\in\KGspace{\Hin}$  is   it   associated    Cauchy  data,  then
$<u^\inn,u^\inn>_{C^\inn}=<{\cal A}u^\inn,{\cal A}u^\inn>_{C^\out}$~.

\vskip 1em
In order to construct ${\cal A}$  from initial Cauchy data
$u^\inn\in\KGspace{\Hin}$  one or
more of the
following obstructions may arise
 when attempting to calculate the corresponding
element $u^\out\in\KGspace{\Hout}$:

(1) There is no compatible solution in any Euclidean neighbourhood of
$\Sigma^\inn$. An illustration of this has been given in  example 1.

(2) Although  a solution in a neighbourhood of $\Sigma^\inn$ exists, 
this solution has a ``natural boundary'' which prevents it
propagating to $\Sigma^\out$.
For example, consider the analytic function $f:{\Bbb D}\mapsto\Cmpx$,
\begin{eqnarray}
f(z)=\sum_{n=0}^\infty \left({z \over R}\right)^{(n!)}
\end{eqnarray}
where ${\Bbb D}\subset\Cmpx$ is the open disk of radius $R$.
This function is infinite for all $z=Re^{p/q},\ p,q\in\Intg$, and
is therefor said to have a natural boundary on $\partial {\Bbb D}$.
It cannot be analytically continued beyond $\partial {\Bbb D}$.
If $1<R<e^\xi$ then $\Psi(\Sigma^\inn)\subset {\Bbb D}$ and
$\Psi(\Sigma^\out)\not\subset {\Bbb D}$, hence 
the state in $\KGspace{\Hin}$ corresponding to $f$ cannot be
propagated to $\Sigma^\out$.

(3) The state propagates to $\Sigma^\out$ but contains a
singularity on $\Sigma^\out$. This may give an  infinite norm for the
state at $C^\out$.

(4) An analytic continuation of the state on $\Sigma^{\inn}$ to
$\Sigma^{\out}$ exists with singularities in $E$. Such singularities
can give rise to non-trivial de-Rham periods and contribute to the
breakdown of unitarity.  In \cite{Jg2} it is suggested that a
resolution to this problem is to excise those domains where such
solutions are singular by attaching extra tubes to the manifold. This
correlates the space of allowable Cauchy data to the global topology
of the manifold. One can then restore unitarity for this particular
initial state by labelling some of the tubes as ``{\bf \inn}'' and the
others as ``{\bf \out}''. This resolution works for any particular
state but cannot be applied to the entire space of states without
removing the entire Euclidean domain.

\vskip 1em


\section{Conclusions}

One of the most striking results of the theorem above is that a
spacetime with a Euclidean domain does not necessarily require a
traditional Cauchy surface in order for one to be able to predict a
global solution of the (scalar) wave equation.  Cauchy data on any
acausal curve may be sufficient.  To effect such a prediction it is
only necessary: (i) that the domain of null dependence of the acausal
curve has a non-trivial intersection with the Euclidean domain, (ii)
the solution in the two domains is connected by ``linear
pointwise-invertible'' junction conditions and (iii) the boundary of
the Euclidean domain is a Cauchy curve for the whole Lorentzian
domain.  If the degenerate metric possesses a spacelike Killing
isometry in the Lorentzian domain then (iii) is automatic.

For example, a two-dimensional cosmology may be modelled with a metric
that is induced by appropriately immersing a paraboloid in Minkowski
3-space. If the Euclidean domain corresponds to a parabolic cap then
the ring of signature change is a Cauchy curve for the prediction of
solutions to the wave equation for the whole paraboloid. However every
acausal segment in the Lorentzian region provides a family of
(disconnected) domains of null determination. If one such domain has a
non-trivial intersection with the ring of signature change then any
global solution satisfying the junction conditions above can be
calculated from just the Cauchy data on the original segment.
However, in general, given arbitrary Cauchy data on such a segment
there may be no such global solution.

Although our results have relied fundamentally on the conformal
structure of the scalar wave equation and the dimensionality of the
background manifold we speculate that many of these features will
persist in theories that lack conformal symmetry, such as the massive
Klein-Gordon Equation in three or more dimensions.  The extension of
our methods would exploit the general theory of Elliptic Partial
Differential
Equations. \cite{Mizo1,Courant1,Hormander1,Saka1,Bitsadze1}

The traditional construction of a local quantum field theory on a
spacetime relies on a number of features that are conspicuously absent
on manifolds with a degenerate metric. Most notably it proves
difficult to construct a space of asymptotic quantum states that can
be connected by a unitary Scattering matrix.  

\begin{figure}
\epsfysize=2.5in
\centerline{\epsffile[20 40 610 310]{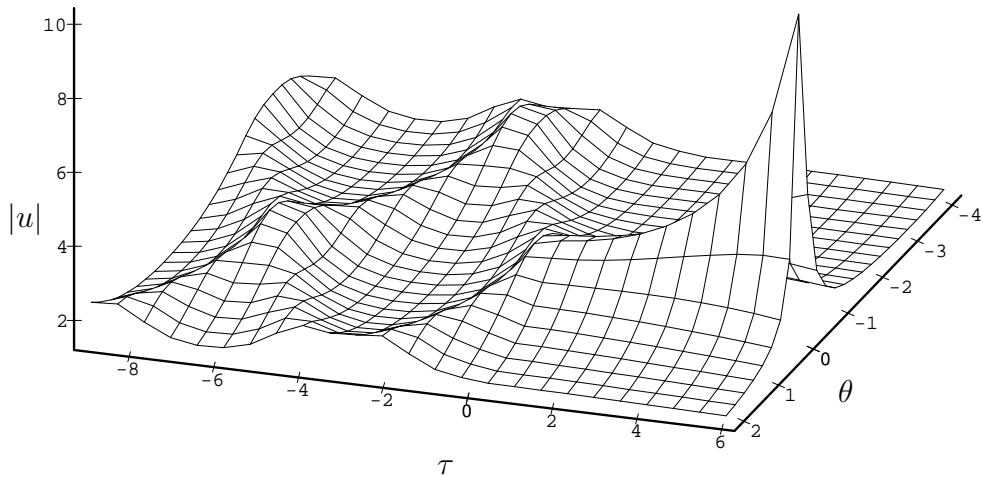}}
\setlength{\unitlength}{1in}%
\begin{picture}(0,0)(-0.5,0)
\put(0.05,1.5){\makebox(0,0)[l]{$|u|$}}
\put(2.3,0.2){\makebox(0,0)[l]{$\tau$}}
\put(4.4,0.6){\makebox(0,0)[l]{$\theta$}}
\end{picture}
\caption{{Evolution with $\tau$ of the modulus of a pair of counter
rotating Lorentzian wavepackets $u$ across a metric degeneracy at
$\tau=0$. The angular coordinate for the cylindrical manifold is
denoted by $0\le\theta<2\pi$. }}
\label{fig_packet}
\end{figure}

Although it is not difficult to find subspaces of bounded Lorentzian
solutions such solutions may become singular when continued to a
Euclidean region via a broad class of matching conditions.  Thus a
well behaved Lorentzian solution can propagate into the future without
attenuation and pass smoothly into a Euclidean domain where it may
rapidly explode.  For example for any $0<r<1$ consider the sum of two
packets
\begin{eqnarray}
u(\tau,\theta)&=& 
\frac1{1-re^{i(\theta+\wh\tau)}} + 
\frac1{1-re^{i(\theta-\wh\tau)}}
\end{eqnarray}
where 
\begin{eqnarray}
\wh\tau=\int_{\tau}^{\tau^\inn}
\left( -g_{\tau\tau}(\tau') \over g_{\theta\theta}(\tau') \right)
^\frac12 d\tau'
\end{eqnarray}
for $(\tau,\theta)\in L^\inn$, and
\begin{eqnarray}
u(\tau,\theta)&=& \frac1{1-z} + \frac{\cnj{z}}{\cnj{z}-r^2}
\end{eqnarray}
for $(\tau,\theta)\in E$ where $z=\Psi(\tau,\theta)$ is given by
(\ref{defPsi}). This is illustrated in FIG. \ref{fig_packet}, with
$r=0.4$, for the cylindrical manifold $M$ above. For $\tau <
\tau^\inn$ the bounded Lorentzian wave packets counter-rotate around
the cylinder unattenuated, until they reach the signature ring at
$\tau^\inn=0$.  This Lorentzian solution is then matched to the
Euclidean one using the junction condition (\ref{natjunc}) but with
the additional constraint that the normal derivatives on either side
of the degeneracy curve vanish.  The solution clearly becomes an
exploding peak as it diffuses into the Euclidean region, becoming
singular before escaping to the Lorentzian domain.

We have
also explicitly demonstrated several other obstructions than can arise
when trying to construct unitary operators in a basis of asymptotic
states.  Although we cannot prove that no such construction is
possible, the results above lead us to strongly suspect that without a
radical departure from traditional methods a local unitary quantum
field theory on a background with a fixed topology and degenerate
metric does not exist.  This conclusion does not necessarily rule out
classical geometries with signature change. A more comprehensive
analysis would consider a coupled field and geometry system with
dynamic topology. Such a quantum geometry would then allow classical
histories describing a manifold with a topology consistent with the
corresponding bounded global field configuration. In the weak-field
semi-classical limit such coupled states might select a
self-consistent classical background geometry with a degenerate metric
upon which one could construct an approximate quantum matter field
description.

\section{ Acknowledgements }

This work has benefited from useful discussions with Charles Wang and
J\"org Schray.  J~G is grateful to the University of Lancaster for a
University Research Studentship and the Royal Society of London for a European
Junior Fellowship. R~W~T is grateful for support from the Human
Capital and Mobility Programme of the EC.

\vskip 1em


\begin{thebibliography}{10}

\bibitem{Geroch1}
R~Geroch.
\newblock Domain of dependence.
\newblock {\em Journal Maths Physics}, 11:437, 1970.

\bibitem{Kundt1}
W~Kundt.
\newblock Non-existence of trouser-worlds.
\newblock {\em Comm. Math. Phys.}, 4:143, 1967.

\bibitem{Gibbons1}
G~W Gibbons.
\newblock The elliptic interpretation of black-holes and quantum-mechanics.
\newblock {\em Nucl. Phys.}, B271:497--508, 1986.

\bibitem{Ellis1}
G~Ellis, A~Sumeruk, D~Coule, and C~Hellaby.
\newblock Change of signature in classical relativity.
\newblock {\em Class. Q. Grav.}, 9:1535--1554, 1992.

\bibitem{Dereli1}
T~Dereli, M~Onder, and R~W Tucker.
\newblock A spinor model for quantum cosmology.
\newblock {\em Phys. Lett. B}, 324:134--140, 1994.

\bibitem{Dereli2}
T~Dereli and R~W Tucker.
\newblock Signature dynamics in general-relativity.
\newblock {\em Class. Quantum Grav.}, 10:365--373, 1993.

\bibitem{Dereli3}
T~Dereli, M~Onder, and R~W Tucker.
\newblock Signature transitions in quantum cosmology.
\newblock {\em Class. Quantum Grav.}, 10:1425--1434, 1993.

\bibitem{Hayward1}
S~A Hayward.
\newblock Signature change in general-relativity.
\newblock {\em Classical And Quantum Gravity}, 9(8):1851--1862, 1992.

\bibitem{Kossowski1}
M~Kossowski and M~Kriele.
\newblock Signature type change and absolute-time in general-relativity.
\newblock {\em Classical Quant Grav}, 10:1157--1164, 1993.

\bibitem{Kossowski2}
M~Kossowski and M~Kriele.
\newblock Smooth and discontinuous signature type change in general-relativity.
\newblock {\em Classical Quant Grav}, 10:2363--2371, 1993.

\bibitem{Kossowski3}
M~Kossowski and M~Kriele.
\newblock Transverse, type changing, pseudo Riemannian metrics and the
  extendibility of geodesics.
\newblock {\em P Roy Soc Lond A Mat}, A444:297, 1994.

\bibitem{Kerner1}
R~Kerner and J~Martin.
\newblock Change of signature and topology in a 5-dimensional cosmological
  model.
\newblock {\em Class. Q. Grav.}, 10:2111--2122, 1993.

\bibitem{Dray1}
T~D Dray, C~A Manogue, and R~W Tucker.
\newblock Scalar field equation in the presence of signature change.
\newblock {\em Phys. Rev. D}, 48:2587--2590, 1993.

\bibitem{Dray2}
T~Dray, C~A Manogue, and R~W Tucker.
\newblock {\em Gen. Rel. Grav.}, 23:967, 1991.

\bibitem{Jg2}
J~Gratus and R~W Tucker.
\newblock Topology change and the propagation of massless fields.
\newblock {\em J. Math. Phys.}, 36(7):3353--3376, 1995.

\bibitem{Alty1}
L~J Alty and C~J Fewster.
\newblock Initial value problems and signature change.
\newblock {\em Class. Q. Grav.} 13:1129-1147, 1996.

\bibitem{Kay1}
B~S Kay.
\newblock {\em Rev. Math. Phys}, 167, 1992.

\bibitem{Spivak1}
M~Spivak.
\newblock {\em A Comprehensive Introduction to Differential Geometry}, volume
  I-VI.
\newblock Publish or Perish, Berkeley, 2nd edition, 1979.

\bibitem{Lang1}
S~Lang.
\newblock {\em Complex Analysis}.
\newblock Springer, 2 edition, 1985.

\bibitem{Mizo1}
S~Mizohata.
\newblock {\em The Theory Of Partial Differential Equations}.
\newblock Cambridge, 1973.

\bibitem{Courant1}
R~Courant and D~Hilbert.
\newblock {\em Methods Of Mathematical Physics}.
\newblock Interscience Pub., 1962.

\bibitem{Saka1}
R~Sakamoto.
\newblock {\em Hyperbolic Boundary Value Problems}.
\newblock Cambridge, 1982.

\bibitem{Fulling1}
S~A Fulling.
\newblock {\em Aspects Of Quantum Field Theory In Curved Space-Time}.
\newblock London Math. Soc. Texts 17. Cambridge, 1989.

\bibitem{Wald1}
R~Wald.
\newblock {\em Quantum Field Theory In Curved Space-Time}.
\newblock Univ. Chicago Press, 1995.

\bibitem{Wald2}
R~Wald.
\newblock {\em General Relativity}.
\newblock Univ. Chicago Press, 1984.

\bibitem{Hormander1}
L~Hormander.
\newblock {\em The Analysis Of Linear Partial Differential Operators}.
\newblock Springer, 1983-85.

\bibitem{Bitsadze1}
A~V Bitsadze.
\newblock {\em Equations of Mixed Type}.
\newblock Pergamon Press, 1964.

\end{thebibliography}

\end{document}